\documentstyle[aps,eqsecnum]{revtex}
\input psfig
\begin{document}
\preprint{NSF-ITP-97-135}
\draft
\title{Resonant Multi-Lead Point-Contact Tunneling
}
\author{C. Nayak$^{1}$, Matthew P.A. Fisher$^{1}$,
A.W.W. Ludwig$^{1,2}$, H.-H. Lin$^{2}$}
\address{
$^{1}$ Institute for Theoretical Physics,
University of California,
Santa Barbara, CA, 93106-4030\\
$^{2}$ Physics Department,
University of California,
Santa Barbara, CA, 93106-4030}
\date{October 28, 1997}
\maketitle

\begin{minipage}{6.0in}

\begin{abstract}
We analyze a model of resonant point-contact tunneling
between multiple
Luttinger liquid leads. The model is a variant
of the multi-channel Kondo model and can be related to
the quantum Brownian motion of a particle on lattices
with $\pi$-flux through each plaquette (in the $3$-lead
case, it is a honeycomb lattice with $\pi$-flux). By comparing the
perturbative and instanton gas expansions, we find a duality
property of the model. At the boundary, this duality
exchanges Neumann and Dirichlet boundary conditions
on the Tomonaga-Luttinger bosons which describe the leads;
in the bulk, it exchanges the `momentum' and `winding' modes
of these bosons. Over a certain range of Luttinger
liquid parameter, $g$, a novel non-trivial intermediate
coupling fixed point controls the low-energy physics.
The finite conductance at this fixed point can be
exactly computed for two special values of $g$.
For larger values of $g$, there is a stable fixed point
at strong coupling which has enhanced conductance resulting from an
analogue of Andreev reflection at the point contact.

\end{abstract}
\vspace{1mm}
\pacs{PACS: 73.23.-b, 71.10.Pm, 73.40.Hm, 73.40.Gk}

\end{minipage}

\section{Introduction}

Despite being a subject of intense interest in recent years,
the study of strongly correlated electron
systems has had a checkered history, primarily
for two reasons. On the one hand, non-perturbative techniques --
of which there are precious few -- are required for their analysis.
At the same time, these systems often exhibit unexpected phenomena,
rendering useless our intuition culled from Fermi liquid
theory and other essentially perturbative problems.
The greatest progress has been made on
one-dimensional systems and, particularly, quantum impurity
problems. In this arena, powerful techniques
such as conformal field theory\cite{affleckludwig}
 and the Bethe {\it ansatz}
\cite{bethe},\cite{FLS}
have led to the discovery of a number of unusual
properties (including spin-charge separation) which
are fundamentally non-perturbative.

In this paper, we analyze
a quantum impurity model which can be physically realized
in a resonant tunneling junction between multiple
quantum wires or quantum Hall edges. Our interest in this
problem is threefold.
First, the results we find -- both intermediate-coupling fixed points
and enhanced conductance due to an analogue of Andreev
reflection at strong-coupling --
are interesting in and of themselves because they truly
are, to use a cliche, exotic.
Second, both the methods
used and the result may shed light
on some of the recurrent themes in the study
of correlated electron systems in which
a single-particle picture is not valid.
In particular, we demonstrate a highly non-trivial
duality that exchanges strong- and weak-coupling.
Recent progress in supersymmetric field theory and
string theory hints at the possibility that such strong-weak coupling
dualties are a common, perhaps even generic, feature of
stongly-coupled field theories. The duality discussed
in this paper has a very rich structure and is one of the
best examples of such a duality in a strongly correlated
electron system. Finally, this model appears to be more generic
and less fine-tuned than many similar ones, which
leads us to hope that our findings could have consequences
for future measurements.

In the next section, we formulate a model describing
several Luttinger liquid leads. Electrons can tunnel at a point-contact
from one of the Luttinger liquids to a resonant state (e.g. a
quantum dot or island); from the resonant state, they can
then proceed and tunnel to another of the Luttinger liquid leads.
A renormalization group analysis shows that when the Luttinger
liquid parameter, $g$, is greater than $1/3$, the tunneling
process is relevant. In section III, following \cite{emerykiv,yikane},
we go to a limit in which we can make an instanton gas
expansion of the strong-coupling limit \cite{duality};
an examination of this limit suggests a strong-weak coupling duality.
This duality leads us to propose the phase diagram of figure 3.
There are three interesting points in this phase diagram
at which we can extract a more detailed understanding of
the physics of this model.
At $g=1$, the electrons in the leads are non-interacting.
If we assume that there is no interaction between the electrons
at the ends of the leads and an electron on the resonant state,
then the problem is a free fermion problem, and can be solved exactly;
the solution is discussed in section IV. If, however,
we assume that there is such an interaction, as
we do for $g\neq 1$\footnote{We need such an interaction
in order to pass to the Toulouse limit, as we discuss
below. The $g=1$ model can be continuously deformed
into the $g\neq 1$ models only when this interaction is
non-vanishing.}, a different fixed point results.
We make a conjecture about the relationship between these
fixed points.
At $g=\sqrt{3}$, the model is self-dual; this property allows
us to deduce the conductance.
Finally, for $g>9$, the strong-coupling
fixed point is stable. At this fixed point, as we explain
in section V, we find
an analog of Andreev reflection, which leads to enhanced
conductance, $G>g$. We also compute charge-transfer
selection rules which elucidate the nature of this fixed
point. We emphasize throughout the place of this
model within the general framework of boundary conformal field
theory and describe the most unusual features -- namely the duality
and the Andreev processes -- from several different points of view.

\section{The Model}

\subsection{The Model and Formalism}

We consider a model in which $N$ leads
are coupled to each other through a resonant state, as
in figure 1. 
One possible realization of this model is
a quantum Hall bar in which quasiparticles or electrons
can tunnel between several edges by first hopping from one
edge to a dot or anti-dot and then hopping from there to another edge.
An alternative implementation of this model is
a resonant tunneling junction between $N$ quantum wires.
The former is more naturally described by the
`unfolded' formalism of figure 2a in which
the leads are described by chiral bosons on an infinite line:
\begin{equation}
S_0  =  \int_{-\infty}^{\infty} d x \int d \tau
 \frac{g}{4\pi}\,{\partial_x}{\phi_i}
({\partial_t} + v{\partial_x}){\phi_i}
\label{lzero}
\end{equation}
$g$ and $v$ are respectively the Luttinger paramater and velocity
of the bosons, which we take, without loss of generality, to
be the same in all leads. The quantum wire problem is
more easily visualized in terms of the `folded'
setup of figue 2b, in which the lead is modelled by a non-chiral
Luttinger liquid on the half-line $x<0$\footnote{The two models
are not quite equivalent since in a quantum wire
or any other non-chiral Luttinger liquid,
the electron creation operator
has spin (i.e. $h-{\bar h}$) $1/2$ and 
scaling dimension $1/2g$ (i.e. $h+{\bar h}$) while the electron
creation operator in a chiral Luttinger liquid (\ref{lzero}) is a
dimension-$1/2g$, spin-$1/2g$ operator.
The two models can be mapped into each other by a transformation
which mixes left- and right-moving modes, but point-contact
tunneling is insensitive to this mixing, so all of our
results apply equally to both the `folded' and `unfolded' model.}:
\begin{equation}
{S_0}  =  {\int_{-\infty}^0}\,dx\,\int\,d\tau\,\,
\frac{1}{8\pi}\,{\partial_\mu}{\varphi_i}
{\partial^\mu}{\varphi_i}
\label{szerof}
\end{equation}
The field, $\varphi$, is taken to have the
periodicity condition\footnote{
The quantity $g$ is related to the
usual compactification radius $r$ of the bosonic string
via $r=2\sqrt{g}$.}
 $\varphi\equiv \varphi + 2\pi(2\sqrt{g})$.
In terms of chiral fields,
$\varphi = {\varphi_R} + {\varphi_L}$. By this `folding'
procedure, we have essentially mapped $\phi(x>0)$ to ${\varphi_L}$,
as depicted in figure 2.

The term which transfers charge to the resonant level is
(in the `unfolded' formalism, the corresponding term is
the same, but with ${\varphi_i}/2\sqrt{g}$ replaced by $\phi$):
\begin{equation}
{S_t} = t\,\int\,d\tau {\sum_j}
\left( {\eta_j}{S^+}{e^{-i{\varphi_j}(0)/2\sqrt{g}}} +
{\eta_j}{S^-} {e^{i{\varphi_j}(0)/2\sqrt{g}}}\right)
\label{ltun}
\end{equation}
Here, we have replaced the charge state of the resonant level
by a spin-$1/2$ degree of freedom. The spin raising and lowering
operators, $S^\pm$, are the creation and annihilation operators for an electron
or quasiparticle on the resonant level. The cocycles, $\eta_i$, must
anticommute, $\{ \eta_i, \eta_j\} = 2 \delta_{ij}$,
 so that the tunneling operators have the correct bosonic
commutation relations. This is true even when (\ref{ltun})
transfers anyonic quasiparticles between the leads,
so this model at $g=3$ describes tunneling between
quantum Hall edges via an anti-dot in the interior
of a Hall droplet at $\nu=1/3$.\footnote{See Appendix 1.}
For $N=3$, the $\eta_i$'s
can be represented by Pauli matrices.
When the leads are decoupled  {\bf ($t=0$)}`, the fields $\varphi_i$
have Neumann boundary conditions at $x=0$;
for $t\neq 0$, some other conformally invariant
boundary condition is dynamically generated in the infrared.

Here, we are assuming that the level is
perfectly resonant and that the different leads
are coupled to this level with
the same hopping strength strength, $t$.
In an experiment, the resonance can be tuned by controlling
one parameter, such as a backgate voltage.
If there are three leads (the simplest case with
a non-trivial phase diagram), then two more parameters
must be tuned to ensure that the hopping strengths
are equal.

The fields $\phi_i$, $\varphi_i$ can be interpreted
in terms of the voltage drops along and between leads.
In the `unfolded' formalism, the field ${\phi_i}$
can be discontinuous across $x=0$,
and this discontinuity, ${\phi_i}(0+)-{\phi_i}(0-)$,
is proportional to the voltage
drop across $x=0$ in the $i^{th}$ lead.
When the leads are decoupled, there is
no voltage drop along the `unfolded' leads,
${\phi_i}(0+)={\phi_i}(0-)$ or,
equivalently, ${\varphi_R}(0)={\varphi_L}(0)$
(Neumann boundary condition).
On the other hand, the voltage drop between leads $i$ and $j$ at the
contact is proportional to ${\varphi_i}(0)-{\varphi_j}(0)$.
In most of the following, we will use the `folded' formalism,
but all of our results can be reinterpreted in the
other language. In the appendix, we discuss the conventions for
these bosonic fields, $\varphi_i$. In particular, we discuss the
mode expansions of these fields and the zero modes,
which play a crucial role in the following analysis.
In terms of the `momentum' zero modes (see appendix),
the Neumann boundary conditions have the effect
of reflecting the zero modes of incoming
states into those of outgoing states,
${P_L^i}={P_R^i}$. When $t\neq 0$, these
momenta are instead shifted, ${P_L^i}={P_R^i}+{Q^i}$.
The allowed shifts, ${Q^i}$, lie on a
lattice which is connected to the problem of
quantum Brownian motion in a periodic potential,
as we will discuss in the next section.

First, however, we note that the Kubo formula for the conductance
(obtained in the usual way, see e.g. \cite{kanefisher}, by introducing
a vector potential, $A$, between
the resonant level and one of the leads, say lead $3$, and
differentiating the partition function with respect to
$A$) takes the following form:
\begin{equation}
G = 2g\,\left( 1 - \frac{|\omega|}{2\pi}
\langle {\varphi_3}(x=0,\omega){\varphi_3}(x=0,-\omega)\rangle \right)
\label{condform}
\end{equation}
When $t=0$, $\varphi_3$ is a free field (Neumann
boundary condition), so$|\omega|
\langle {\varphi_3}(x=0,\omega){\varphi_3}(x=0,\omega)\rangle = 2\pi$
and therefore $G=0$, as we would expect since the leads are decoupled.

\subsection{The Toulouse Limit and Quantum Brownian Motion}

Let us focus, for the moment, on the case $N=3$. The generalization
to arbitrary $N$ is straightforward.
We rewrite (\ref{lzero}) and
(\ref{ltun}) as:
\begin{equation}
{\cal L} = \frac{1}{8\pi}\,{\partial_\mu}{\varphi_i}
{\partial^\mu}{\varphi_i} \,\, + \,\,
t\,\delta(x)\,
{\sum_j}\left( {\eta_j}{S^+}
{e^{-i ( {{\bf R}_\parallel^j} \cdot {\bf \varphi}\,+
\,{{\bf R}_\perp} \cdot {\bf \varphi}) }}
\,+\, {\eta_j}{S^-} {e^{i ( {{\bf R}_\parallel^j}\cdot {\bf \varphi}\,+\,
{{\bf R}_\perp} \cdot {\bf \varphi} ) }}\right)
\label{linter}
\end{equation}
where ${\bf \varphi} = ({\varphi_1},{\varphi_2},{\varphi_3})$ and
${{\bf R}_\parallel^1} = \frac{1}{6\sqrt{g}} \, (2,-1,-1)$,
${{\bf R}_\parallel^2} = \frac{1}{6\sqrt{g}} \, (-1,2,-1)$,
${{\bf R}_\parallel^3} = \frac{1}{6\sqrt{g}} \, (-1,-1,2)$,
${{\bf R}_\perp} = \frac{1}{3\sqrt{2g}} \, (1,1,1)$.
With this notation, we have anticipated the mapping to the
problem of quantum Brownian motion on a honeycomb lattice
with lattice vectors ${{\bf R}_\parallel^1}, {{\bf R}_\parallel^2},
{{\bf R}_\parallel^3}$ and $S_z$ keeping track of the sublattice.
There is one step left before such
an identification can be complete, namely decoupling
${{\bf R}_\perp}\cdot{\bf \varphi}$ by going to the Toulouse
limit as in \cite{emerykiv,yikane}. To do this, we modify
the Hamiltonian related to (\ref{linter})
by
\begin{equation}
{\cal H} \rightarrow {\cal H} + 
{{\tilde t}_z}\,\delta(x)\, ({ \sum_i}\, {\Pi}_i )
\,{S_z}/2\sqrt{g},
\label{Toulouse}
\end{equation}
where $\Pi$ is the momentum conjugate
to $\varphi$.
The added term is an interaction between the charge of the
resonant level and the charge density of the lead at the point
contact. As a result of this term, the modified Lagrangian
describes an interacting system even at $g=1$ although in this
case the
interaction takes place only at $x=0$. It is natural to assume
that this term does not affect the low-energy physics
of (\ref{linter}) or, in other words, that (\ref{linter}) and
the modified Lagrangian flow to the same infrared fixed point.
This does not, appear to be the case at $g=1$, but
this might be special to $g=1$.
The advantage of adding such a term to the Lagrangian is
that we can now \cite{emerykiv,yikane} perform a canonical
transformation
\footnote{To be more
rigorous, we should use $U = {e^{i{{\tilde t}_z} {\sum_i}\,
{\varphi_i}(0)\,{S_z}/2\sqrt{g} }}
\,\,{e^{-i{{\tilde t}_z} {\sum_i}\,{\varphi_i}(-\infty)\,{S_z}/2\sqrt{g} }}$,
which is overall charge neutral since only integer
charges can be added to the systen. The second exponential
compensates the fractional charge added at $0$ by removing
an equal amount at $-\infty$.}
 generated by
$U = {e^{i{{\tilde t}_z} {\sum_i}\,
{\varphi_i}(0)\,{S_z}/2\sqrt{g} }}$. This
has the effect of simultaneously removing the term which we just
added and removing the ${{\bf R}_\perp}\cdot{\bf \varphi}$ terms
from the exponentials in the tunneling Lagrangian if
we choose ${{\tilde t}_z} = 1/N$. This
leaves us, finally, with the Lagrangian:
\begin{equation}
{\cal L} = \frac{1}{8\pi}\,{\partial_\mu}{k_i}
{\partial^\mu}{k_i} \,\, + \,\,
t\,\delta(x)\,
{\sum_j}\left( {\eta_j}{S^+}
{e^{-i{{\bf R}^j} \cdot {\bf k} }}
\,+\, {\eta_j}{S^-} {e^{i{{\bf R}^j} \cdot {\bf k} }}\right)
\label{lqbm}
\end{equation}
where
\begin{equation}
\label{kdef}
{\bf k}~=~(\frac{1}{\sqrt{2}}(-{\varphi_1}+{\varphi_2}),
\frac{1}{\sqrt{6}}(-{\varphi_1}-{\varphi_2}+2{\varphi_3}),
\frac{1}{\sqrt{3}}({\varphi_1}+{\varphi_2}+{\varphi_3}))
\end{equation}
and 
\begin{eqnarray}
\label{Rdef}
{{\bf R}^1} &=& \sqrt{\frac{1}{3g}}\,(-\sqrt{3}/2,-1/2,0)\cr
{{\bf R}^2} &=& \sqrt{\frac{1}{3g}}\,(\sqrt{3}/2,-1/2,0)\cr
{{\bf R}^3} &=& \sqrt{\frac{1}{3g}}\,(0,1,0)
\end{eqnarray}
$\partial k_x$ and $\partial k_y$ are the Cartan
generators of an $SU(3)$ that `rotate' the leads
(which is a symmetry of the free Lagrangian at certain
special points such as $g=1, 1/2$).
Yi and Kane \cite{yikane} showed that the 3-channel Kondo problem
is one of a class of models (namely the $g=1/2$ point) which may be
formulated as the quantum Brownian motion of a particle
on a honeycomb lattice. In (\ref{lqbm}), we have almost
the same problem. The crucial difference is the presence of
the $\eta_i$'s which results in a $\pi$ flux through each plaquette
of the honeycomb lattice. This may be seen by considering
the amplitude for a circuit around a plaquette, which involves
the product ${\eta_1}{\eta_2}{\eta_3}{\eta_1}{\eta_2}{\eta_3}=-1$.

The RG equation for $t$ may be obtained from the scaling
dimension of the field ${e^{i{{\bf R}^i} \cdot {\bf k} }}$,
which is ${\left|{{\bf R}^i}\right|^2} = 1/3g$:
\begin{equation}
\frac{dt}{dl} = \left(1 - \frac{1}{3g}\right)\,t + \ldots
\label{rgt}
\end{equation}
Hence, for $g<1/3$, $t$ flows to zero in the infrared and the
leads are decoupled. For $g>1/3$, $t$ grows with decreasing
energy scales. The upshot of this growth will be analyzed in
the next section using a duality property of this model.

The partition function may be expanded perturbatively
in powers of $t$:
\begin{equation}
Z = {\sum_n}\,{\sum_{\{{l_j}\}}}\frac{t^n}{n!}\,\int{d\tau_1}\ldots{d\tau_n}\,
\delta\left({\sum_j}{\epsilon_j}{{\bf R}_{l_j}}\right)
{\rm exp}\left({\sum_{i>j}} \,{\epsilon_i}{\epsilon_j}\,
{{\bf R}_{l_i}}\cdot{{\bf R}_{l_j}}\,\ln|{\tau_i}-{\tau_j}|\,+\,
i\pi\,\theta({\tau_i}-{\tau_j})\,
\left(1-{\delta_{{l_i}{l_j}}}\right)\right)
\end{equation}
${l_i}=1,2,3$, $\epsilon_i =\pm 1$ and the $\epsilon_i$'s
must alternate chronologically. If we ignore the
second term in the exponential, this is the partition
function (at $g=1/2$) of the 3-channel Kondo model.
It is a two-component Coulomb gas.
The second term gives a minus sign whenever the
order of two unlike hops is exchanged, thereby
implementing the $\pi$-flux.

\subsection{An Auxiliary Model}

We will also consider a simpler model (which is discussed
in \cite{yikane}) for the purposes of comparison with and illumination of
the resonant tunneling model described above.
This model can be analyzed without going
to a Toulouse limit, and it exhibits Andreev reflection at a
strong coupling fixed point and a duality property
with a straightforward interpretation. This
instills us with more confidence that these properties
of (\ref{lqbm}) are generic and are not particular to
the Toulouse limit. It is defined by
\begin{equation}
{\cal L} = \frac{1}{4\pi}\,{\partial_x}{\phi_i}
({\partial_\tau} + i{\partial_x}){\phi_i} \,\, + \,\,
t\,\delta(x)\,
{\sum_{i\neq j=1,2,3}}\left(
{e^{-i ({\phi_i} - {\phi_j})/\sqrt{2g} }}
\,+\,  {e^{i ({\phi_i} - {\phi_j})/\sqrt{2g} }}\right)
\label{laux}
\end{equation}
This is a model of quantum Brownian motion on a
triangular lattice. In fact, this is the same triangular lattice
which is the underlying Bravais lattice of the above
honeycomb lattice, as may be seen by writing the Lagrangian
(\ref{laux}) as ${({\partial_\mu} {\bf k})^2} + t\delta(x)
{\sum_j} {e^{{\bf k}\cdot{{\bf R}_j}}} + {\rm h.c.}$
with ${\bf k}$ as in (\ref{kdef}) and ${{\bf R}_j}$
given by (\ref{Rdef}) with the first and second components
interchanged.
At $g=1$, (\ref{laux}) has a fermionic
representation:
\begin{equation}
{\cal L} = {\psi^\dagger_i}\left({\partial_\tau}+i{\partial_x}\right){\psi_i}
\,+\,t\,\delta(x)\,{\epsilon^{ijk}}{\eta_k}{\psi^\dagger_i}{\psi_j}
\end{equation}
This is not a free fermion problem because the fermion
interacts with a spin-$1/2$ degreee of freedom, $\eta$,
which is present to give the correct commutation relations,
as in (\ref{ltun}).
This model is actually a generalized multi-channel
Kondo model in which the conduction electrons transform in an $SU(2)$
triplet. The infrared fixed point can be solved for exactly\cite{sengupta}
in complete analogy with the methods employed
in the ordinary multi-channel Kondo model\cite{affleckludwig}.
Interestingly, it is related to the ordinary  four-channel, spin $1/2$
Kondo fixed point. According to
\cite{yikane}, the model flows to strong coupling
at $g=1$. Hence, the fixed point of \cite{sengupta}
is an example of the Andreev reflection phenomenon
which we discuss below.
The advantage of this model lies in the fact that
there are no complications related to the Toulouse
limit, as there are in (\ref{lqbm}). It is particularly
simple from the point of view of duality.

\section{Duality}

If $g>1/3$, $t$ is a relevant coupling, so an initially
small $t$ grows in the infrared. When $t$ is large, the
interaction term, $t\,\delta(x)\,
{\sum_i}\left( {\eta_j}{S^+}
{e^{-i{{\bf R}^i} \cdot {\bf k} }}
\,+\, {\eta_j}{S^-} {e^{i{{\bf R}^i} \cdot {\bf k} }}\right)$
will be dominant and, in a semi-classical analysis,
$k$ will be localized at one of its minima.
These minima are just the minima of the energy bands
of a particle on a tight-binding honeycomb lattice with
$\pi$ flux per plaquette. There are four such energy bands
since the $\pi$ flux doubles the unit cell and since the
honeycomb lattice, to begin with, is a triangular lattice with a two
site basis. (We represent $\eta_j$ by Pauli
matrices  $\tau_j$.)
They correspond to the four possible $S_z$ and
$\tau_3$ quantum numbers. At low energies, $k$ will be in one
of the minima of the lowest band. These also
form a honeycomb lattice; the lattice displacements -- i.e.
the analogs of the ${\bf R}_i$'s -- on this honeycomb
lattice are
\begin{eqnarray}
\label{qs}
{{\bf K}_1} &=& \frac{\sqrt{g}}{3}(0,1)\cr
{{\bf K}_2} &=& \frac{\sqrt{g}}{3}(\sqrt{3}/2,-1/2)\cr
{{\bf K}_3} &=& \frac{\sqrt{g}}{3}(-\sqrt{3}/2,-1/2)
\end{eqnarray}

The partition function can be approximated by an instanton
gas in which the instantons are solutions of the
Euclidean equations of motion in which $k$ tunnels between
different minima. As usual in this class of problems
\cite{duality}, the instanton gas expansion
can be formulated as a Coulomb gas.
There is an additional subtlety here, however:
there is a Berry's phase associated with the instanton
solutions. Details will be given in an appendix;
here we merely sketch the derivation.
Note that the minima of the lowest band
surround a point at which the two lowest bands touch.
The Berry's phase will be the same for any path surrounding
this point, so we consider a path that is very
close to this point. For such paths, ${\sum_j}\left( {\eta_j}{S^+}
{e^{-i{{\bf R}^j} \cdot {\bf k} }}
\,+\, {\eta_j}{S^-} {e^{i{{\bf R}^j} \cdot {\bf k} }}\right)$
can essentially be approximated by
$-{\delta k_x}{\sigma_z}-{\delta k_y}{\sigma_x}$.
Here, the four energy bands, acted on by $\eta\otimes S$,
are reduced to the two-dimensional subspace of the
two lowest bands, acted on by the $\sigma$'s. $\delta {k_x}$,
${\delta k_y}$ are $k_x$, $k_y$ measured from the contact
point of the two bands. As $\delta {\bf k}$ traces out a path
around $0$, the spin $\sigma$
rotates by $2\pi$ and therefore accrues a Berry phase of $\pi$.
Hence, the Coulomb gas defined by the instanton expansion
is a Coulomb gas with phases. In fact, it is of precisely
the same variety as that defined by the perturbative expansion
of (\ref{lqbm})! More concretely, the instanton -- or strong-coupling --
expansion of (\ref{lqbm}) is equal to the perturbative --
or weak-coupling -- expansion of
\begin{equation}
{{\cal L}_D} = \frac{1}{8\pi}\,{\partial_\mu}{r_i}
{\partial^\mu}{r_i} \,\, + \,\,
v\,\delta(x)\,
{\sum_i}\left( {\eta_j}{S^+}
{e^{-i{{\bf K}^j} \cdot {\bf r} }}
\,+\, {\eta_j}{S^-} {e^{i{{\bf K}^j} \cdot {\bf r} }}\right)
\label{ldual}
\end{equation}
(Here $\bf r$ is the field dual to the field ${\bf k}$
of (\ref{kdef}) in the usual way, as reviewed in Appendix C.)
The $v\rightarrow 0$ limit of (\ref{ldual}) is equivalent
to the $t\rightarrow\infty$ limit of (\ref{lqbm}) and, conversely,
the $v\rightarrow\infty$ limit is equivalent to the
$t\rightarrow 0$ limit. In effect,  the duality
exchanges $g\rightarrow 3/g$. For small $v$, we can obtain
the RG equation for $v$ just as we did for $t$ above:
\begin{equation}
\frac{dv}{d l} = \left(1 - \frac{g}{9}\right)\,v \,+ \ldots
\label{rgv}
\end{equation}

Combining (\ref{rgt}) and (\ref{rgv}), we find that the
$t=0$ limit is stable for $g<1/3$ while the $t=\infty$
limit is stable for $g>9$. In the former limit,
the fields ${k_1}, {k_2}$ have Neumann boundary conditions
at $x=0$ ($k_3$ is decoupled, so it always has Neumann boundary
conditions), while the $r_i$ have Dirichlet boundary
conditions. In the latter limit, the situation is
reversed. For $1/3<g<9$, both limits
are unstable and we expect a stable fixed point at intermediate
coupling or, in other words, a non-trivial conformally
invariant boundary condition. The situation is summarized by figure 2.
There are two intermediate coupling fixed points at which
we can calculate the conductance exactly:

(a) At $g=1$, where a free fermion formulation
is available for ${{\tilde t}_z} = 0$. We do not
believe that the ${{\tilde t}_z} = 0$ model has
the same physics as the ${{\tilde t}_z} = 1/N$
model, but it is instructive to compare the
two cases.

(b) At $g=\sqrt{3}$, the model is self-dual. It
may be shown \cite{duality} that the duality exchanges
$\frac{|\omega|}{2\pi}\langle {k_x}{k_x}\rangle \rightarrow 1
- \frac{|\omega|}{2\pi}\langle {r_x}{r_x}\rangle$
(and the same for $k_y$) as we discuss in
an appendix.
 At the self-dual point,
$\langle {k_x}{k_x}\rangle=\langle {r_x}{r_x}\rangle$,
so $G=g(2/3)=2/\sqrt{3}$.

\noindent{We will also discuss at length the conductance at:}

(c) The strong-coupling fixed point, which is
stable for $g>9$.

First, however, we will make a few more comments on the duality
between (\ref{lqbm}) and (\ref{ldual}).
One point which should be emphasized is that the duality
is only approximate. It is strictly a duality between the
instanton gas expansion of (\ref{lqbm}) and the perturbative
expansion of (\ref{ldual}) (and vice versa). In the asymptotic
low-energy limit, the instanton gas expansion of (\ref{lqbm}) is
the dominant contribution to the partition function when
$t$ is large, but at finite energy there are corrections.
If we were to attempt to formulate an exact duality, these
corrections would be manifested by the presence of
a presumably infinite number of
additional
irrelevant terms in (\ref{ldual})\cite{FLS}.

The perturbative expansion of (\ref{lqbm})
is an expansion in current-generating charge transfer
events while the instanton gas is an expansion
in voltage-generating phase slips (in (\ref{ldual}),
the roles are reversed). This formulation of the duality
concentrates on the values of the
fields at the point contact.
A related but alternative way of
understanding this
duality arises from the natural notion of duality
inherent in the bulk (i.e. the duality of closed
strings with toroidal compactification).
Let us first look at the simpler model (\ref{laux}).
Following the same steps which led to (\ref{ldual}),
we see that (\ref{laux}) is dual to a theory described
by the same Lagrangian (\ref{laux}), but with the replacement
$g\rightarrow 3/4g$ \cite{yikane}. Let us consider
the finite temperature partition function of this model
in a finite-size system of length $L$, with Neumann
boundary condition at $x=L$ and the interaction at $x=0$,
as in fig. 4a. This partition function can also be viewed
(by turning it on its side)
as the closed string amplitude for propagation
between the dynamical boundary state at $x=0$
and the Neumann boundary state at $x=L$, as in fig. 4b.
The closed string states are specified by their momenta,
winding numbers, and oscillator mode occupancies (see Appendix
C for a brief  summary).
The allowed momenta, ${\bf P}$, for the fields ${\bf k}=
(\frac{1}{\sqrt{2}}(-{\varphi_1}+{\varphi_2}),
\frac{1}{\sqrt{6}}(-{\varphi_1}-{\varphi_2}+2{\varphi_3}))$
are determined by the condition that the operator
$e^{i{\bf P}\cdot{\bf k}}$ be well-defined
under the identification
${\varphi_i}\equiv{\varphi_i}+2\pi\sqrt{2g}$; the momenta form
a triangular lattice with lattice constant
$\sqrt{1/g}$.\footnote{As usual, we ignore
${k_z} = \frac{1}{\sqrt{3}}({\varphi_1}+{\varphi_2}+{\varphi_3})$,
which decouples.}
The winding numbers, ${\bf W}$ are the set of identifications,
${\bf k} \equiv {\bf k} + {\bf W}$; they form a triangular
lattice with lattice constant $2\sqrt{g/3}$.
There are two dual descriptions which result from
exchanging the momenta and winding modes. This is
precisely the same duality between triangular lattices
which exchanges
the strong and weak coupling limits of (\ref{laux}).
The model (\ref{lqbm}) can be embedded within this picture.
The only additional structure is that the displacements
on the triangular lattice (i.e. charge transfers or phase slips)
are split into pairs of displacements on the honeycomb lattice
in both the original and dual theories. Yet another interpretation
in terms of $S$-matrix selection rules will be dicussed
in the context of the Dirichlet boundary condition
\cite{chamonfradkin}.

\section{Solution at $g=1$}

At $g=1$, the model defined by (\ref{lzero}) and (\ref{ltun})
has the free fermion representation (in particular,
with ${{\tilde t}_z} = 0$):
\begin{equation}
{\cal L} = {\psi^\dagger_i}\left({\partial_\tau}+i{\partial_x}\right){\psi_i}
\,+\, {d^\dagger}{\partial_\tau}d
\,+\,t\,\delta(x)\,
{\sum_i}\,{\psi^\dagger_i}d\,+\,{\psi_i}{d^\dagger}
\end{equation}
The creation and annihilation operators of charge on the
resonant state are denoted by ${d^\dagger}$, $d$
rather than $S^\pm$, and $\{d,\psi\}=\{d,{\psi^\dagger}\} = 0$,
$\{d,{d^\dagger}\} = 1$. This free fermion problem can be
solved exactly. The equations of motion for $\psi$ and $d$ are:
\begin{eqnarray}
{\partial_\tau}{\psi_i}(x) = {\partial_x}{\psi_i}(x) + t\,d\,\delta(x)\cr
{\partial_\tau}d = t\,{\sum_i}{\psi_i}(0)
\end{eqnarray}
Integrating the first equation between $x=-\epsilon$ and $x=\epsilon$
and Fourier transforming, we find
\begin{eqnarray}
{\psi_i}(\omega,0+)-{\psi_i}(\omega,0-) = - it d(\omega)\cr
\omega d(\omega) = t {\sum_i}{\psi_i}(\omega)
\end{eqnarray}
In the second equation,
${\psi_i} = ({\psi_i}(0+)+{\psi_i}(0-))/2$. From these $N+1$
equations, we can extract ${\psi_i}(\omega,0+)$
and $d(\omega)$ in terms of ${\psi_i}(\omega,0-)$. The solution
may be summarized by the $S$-matrix, ${\psi_i}(\omega,0+) =
{S_{ij}}\,{\psi_j}(\omega,0-)$, where:
\begin{eqnarray}
{S_{ii}}&=& \frac{N-2}{N}\cr
{S_{ij}}&=& \frac{-2}{N}, \,\,{\rm for\,\, i\neq j}
\end{eqnarray}
The resulting conductance is
\begin{equation}
{G_{\rm free\,fermion}} = (N-1)\,{\left(\frac{2}{N}\right)^2}
\end{equation}
which, for $3$ leads is $G=8/9$. 
It is somehwat remarkable that a free fermion problem
could be an interediate-coupling fixed point with
a non-trivial conductance.
However, this is the maximal possible conductance consistent
with unitarity and permutation symmetry for a $3$ lead free fermion problem.
In other words, if we assume that 
\begin{eqnarray}
{S_{ii}}&=& r\cr
{S_{ij}}&=& t, \,\,{\rm for\,\, i\neq j}
\end{eqnarray}
then unitarity, ${S_{ij}}{{S^*}_{kj}} = {\delta_{ik}}$,
imposes the constraint $|r| \geq 1/3$,
and, hence, $G \leq 8/9$.
In the next section,
we will discuss even larger conductances and
the physics behind them.

First, however, we will comment on the relationship
between the ${{\tilde t}_z} = 0$ and ${{\tilde t}_z} = 1/N$
fixed points. We do not believe that they are the same
for two reasons. First, we expect $G/g$ to be
non-decreasing as $g$ is increased. While
$G(g=1, {{\tilde t}_z} = 0) < G(g=\sqrt{3}, {{\tilde t}_z} = 1/3)$,
$G(g=1, {{\tilde t}_z} = 0) >
G(g=\sqrt{3}, {{\tilde t}_z} = 1/3)/\sqrt{3}$.
Hence, we expect that 
$G(g=1, {{\tilde t}_z} = 1/3) <
G(g=\sqrt{3}, {{\tilde t}_z} = 1/3)/\sqrt{3}
< G(g=1, {{\tilde t}_z} = 0)$.
An additional point for consideration is
that a small ${{\tilde t}_z}$ is an irrelevant
perturbation at the ${{\tilde t}_z} = 0$ fixed point,
as may be seen by direct calculation. Similarly,
a small deviation of ${{\tilde t}_z}$ from $1/3$
is irrelevant at the ${{\tilde t}_z} = 1/3$ fixed point,
as may be shown perturbatively for $g\rightarrow 1/3$;
it is reasonable to assume that this is true even at $g=1$.
Hence, it is plausible that the ${{\tilde t}_z} = 0$ fixed point
described above lies out of the plane of the phase
diagram of figure 3 with an unstable fixed point separating it
from the ${{\tilde t}_z} = 1/3$ fixed point.

\section{Dirichlet Boundary Conditions and Andreev Reflection}

A remarkable feature of this model reveals itself when
we consider the conductance at the $t=\infty$ fixed point,
which is stable for $g>9$ (and for $g>1$ in the auxiliary
model (\ref{laux})). At this fixed point,
${k_1} = \frac{1}{\sqrt{2}}(-{\varphi_1}+{\varphi_2})$ and
${k_2} = \frac{1}{\sqrt{6}}(-{\varphi_1}-{\varphi_2}+2{\varphi_3})$
have Dirichlet boundary conditions at $x=0$, while
${k_3} = \frac{1}{\sqrt{3}}({\varphi_1}+{\varphi_2}+{\varphi_3})$
has Neumann boundary condition. As a result,
\begin{eqnarray}
\frac{|\omega|}{2\pi}
\langle {\varphi_3}(x=0,\omega){\varphi_3}(x=0,\omega)\rangle
&=& \frac{2}{3}\,
\frac{|\omega|}{2\pi}
\langle {k_1}(0,\omega){k_1}(0,\omega)\rangle
\,\,+\,\, \frac{1}{3}\,
\frac{|\omega|}{2\pi}
\langle {k_3}(0,\omega){k_3}(0,\omega)\rangle\cr
&=&  \frac{2}{3}(0) + \frac{1}{3}(1) = \frac{1}{3}
\end{eqnarray}
where the second equality follows from the respective
Dirichlet and Neumann boundary conditions of $k_1$ and
$k_3$. Hence, from (\ref{condform}) we have:
\begin{equation}
{G_3^{max}} = \frac{4}{3}g
\end{equation}
This is an astonishing result, since it implies that the
conductance is greater than `perfect' conductance,
$G=g$.\footnote{The scrupulous reader might worry that this surprising
finding is due entirely
to the Toulouse limit and is therefore incorrect.
However, since the same conductance is found for (\ref{laux})
(which does not involve a Toulouse limit)
at its strong-coupling fixed point, we believe that
this result is robust.}
We interpret this as the signature of {\it Andreev
reflection}: the conductance is greater than its naive
maximum value because a hole is backscattered at the point
contact. Before pursuing this point further, let us
note that for general $N$, the corresponding formula
for the conductance at the strong-coupling fixed point
is:
\begin{equation}
{G_N^{max}} = g\left(2 - \frac{2}{N}\right)
\end{equation}
For $N=2$, the maximum conductance is $G=g$, the naive value.
For $N>2$, the maximum conductance is greater than this value,
saturating at $G=2g$ in the $N\rightarrow\infty$ limit.

Why do we say that the enhanced conductance is due
to Andreev reflection? In (\ref{condform}),
$2 - 2\frac{|\omega|}{2\pi}\langle{\varphi_3}{\varphi_3} \rangle$
is, essentially, the transmitted fraction of the
incoming current; $2\frac{|\omega|}{2\pi}
\langle{\varphi_3}{\varphi_3} \rangle-1$
is the reflected fraction. At $g=1$, where the leads
have a free fermion description, transmission, $t$,
and reflection, $r$, coefficients can be defined;
$2\frac{|\omega|}{2\pi}\langle{\varphi_3}{\varphi_3}
\rangle - 1 = {|r|^2}$.
By charge conservation we also have $ (N-1) |t|^2 + |r|^2 =1 $.
$G>g$ precisely because
the reflection coefficient is {\it negative}. In other words,
the reflected current is a {\it negative} fraction of
the incoming current -- i.e. it is a current of holes
rather than electrons.

Physically, the Dirichlet
boundary condition corresponds to the limit
in which there is no voltage difference between the
different leads. For $N-1>1$ only a fraction of the current
which leaves one lead enters any one of the other
$N-1$ leads. Without Andreev scattering, this would lead
to a voltage drop between the leads, but Andreev
processes offset this voltage.
An alternative perspective on the multichannel
Dirichlet boundary condition is reminiscent
of the situation explored in \cite{chamonfradkin}.
Suppose we view $N-1$ of the leads as a single,
aggregate lead described by a single charge boson with
${g_{aggr}} = g(N-1)$. Then, tunneling between
the remaining lead and the aggregate lead is
precisely the problem of tunneling between
dissimilar Luttinger liquids considered in
\cite{chamonfradkin}.
This problem can be transformed to one with
two identical Luttinger liquids with
$1/{g_{eff}} = (1/g + 1/{g_{aggr}})/2 = N/2(N-1)g$.
For such a problem, it is not surprising that
the maximal conductance is
${G_{max}} = {g_{eff}} = 2g(N-1)/N$.

Yet another means of characterizing the Dirichlet boundary
condition is by $S$-matrix selection rules for
soliton scattering at the junction. Following
\cite{chamonfradkin}, we obtain these
by rewriting the chiral fields $\phi_1$, $\phi_2$, $\phi_3$ in
terms of the dual fields ${\tilde k}_x$, ${\tilde k}_y$,
which are free fields at the strong coupling
Dirichlet boundary condition fixed point.
Working in the chiral (`unfolded') notation,
where ${k_x} = ({\phi_1}-{\phi_2})/\sqrt{2},
{k_y}=(-{\phi_1}-{\phi_2}+2{\phi_3})/\sqrt{6}$,
we can define dual free fields via
\begin{eqnarray}
{k_x} &=& {{\tilde k}_x}\,\theta(-x)\,-\,{{\tilde k}_x}\,\theta(x)\cr
{k_y} &=& {{\tilde k}_y}\,\theta(-x)\,-\,{{\tilde k}_y}\,\theta(x)
\end{eqnarray}
These dual fields can be identified with ${r_x}$ and ${r_y}$
which occur in the unfolded form of the dual action (\ref{ldual}).
Now, we can express the $\phi_i$'s in terms of
the free fields ${\tilde k}_x$, ${\tilde k}_y$,
which are just $r_x$, $r_y$.
This allows us to calculate the matrix elements
\begin{equation}
\left\langle\,{e^{-i{\sum_j}{q_j^{\rm out}}{\phi_j}(x=\infty)/\sqrt{2g}}}\,
{e^{i{\sum_j}{q_j^{\rm in}}{\phi_j}(x=-\infty)/\sqrt{2g}}}
\right\rangle
\label{selrulemat}
\end{equation}
The operators ${e^{\mp i{q_j^{\rm in,out}}{\phi_j}(\pm\infty)/\sqrt{2g}}}$
create or destroy states with well-defined charges
in the leads; the matrix elements (\ref{selrulemat})
are proportional to the $S$-matrix elements between
these different charge sectors. For generic
${q_j^{\rm in,out}}$, (\ref{selrulemat}) will vanish, which means
that there is no scattering between these charge sectors
in the strong coupling (Dirichlet boundary condition)
limit. (\ref{selrulemat}) will be non-vanishing only if
the correlation function is charge neutral
for each of the free fields $r_x$, $r_y$, $k_z$. Or,
in other words, if
\begin{eqnarray}
{q_1^{\rm in}} + {q_2^{\rm in}} + {q_3^{\rm in}}
&=& {q_1^{\rm out}} + {q_2^{\rm out}} + {q_3^{\rm out}}\cr
{q_1^{\rm in}} - {q_2^{\rm in}}
&=& -\,\left({q_1^{\rm out}} - {q_2^{\rm out}}\right)\cr
- {q_1^{\rm in}} - {q_2^{\rm in}} + 2{q_3^{\rm in}}
&=& -\,\left(-{q_1^{\rm out}} - {q_2^{\rm out}} + 2{q_3^{\rm out}}\right)
\end{eqnarray}

Solving for the charges of the `out'-states, one finds
that the charge transfers lie on a honeycomb lattice,
\begin{equation}
\Delta {\vec q}\equiv
\pmatrix{q_1^{in}\cr q_2^{in }\cr q_3^{in}\cr}
-\pmatrix{q_1^{out}\cr q_2^{out}\cr q_3^{out}\cr}
=
{2 \over 3} \
\biggl [
\pmatrix{2\cr -1 \cr -1}
q_1^{in}
+
\pmatrix{-1\cr 2 \cr -1}
q_2^{in}
+
\pmatrix{-1\cr -1 \cr 2}
q_3^{in}
\biggl ]
\label{chargetransfersAndreev}
\end{equation}

Note that, for general `in' states which carry in each lead
multiples of the unit of charge, the charges of the
`out'-state  in the individul leads are in general no longer
multiples of the unit charge. This is a phenomenon
analogous to
the $N \geq 3$ flavor Callan-Rubakov effect(\cite{CRE}).  
In fact, the `auxiliary model' at $g=1$, discussed  at the end of
Section II,  is an example where this situation occurs
at an infrared fixed point
for free electron leads.
For example,
\begin{equation}
\left\langle\,{e^{-i {\phi_j}(x=\infty)/\sqrt{2g}}}
{e^{i{\phi_1}(-\infty)/\sqrt{2g}}}\,\right\rangle = 0
\end{equation}
for $j=1,2,3$. In other words, a unit of charge
cannot be transferred from one lead to another
or even reflected by the junction! On the other
hand,
\begin{equation}
\left\langle\,
{e^{i{\phi_1}(x=\infty)/\sqrt{2g}}}
{e^{-2i{\phi_2}(\infty)/\sqrt{2g}}}
{e^{-2i{\phi_3}(\infty)/\sqrt{2g}}}
{e^{3i{\phi_1}(-\infty)/\sqrt{2g}}}\,\right\rangle \neq 0
\end{equation}
This is a clear illustration of the Andreev reflection
property of the Dirichlet boundary condition for $N>2$.
Three incoming electrons in lead $1$ are scattered
into $2$ electrons into each of leads $2$ and $3$
and an Andreev reflected hole in lead $1$.

\begin{acknowledgements}
This work supported in part by NSF grant PHY94-07194 at
ITP-UCSB. C.N. would like to thank C. de C. Chamon for
discussions. C.N. and A.W.W.L. thank E. Fradkin
for an explanation of his work. A.W.W.L. thanks the A.P.
Sloan Foundation for financial support.
\end{acknowledgements}

\appendix

\section{Commutation Relations for the Klein Factors}

We need to introduce the all-important
Klein factors, $\eta_i$, because we would
like to treat the fields $\varphi_i$
as independent bosons which commute with each other.
Since the underlying electron or quasiparticle
operators are mutually fermionic or even anyonic,
Klein factors must be introduced to compensate.

We begin with the commutation relations for
the chiral version of the
tunneling operators, ${T_{ij}}={e^{i{\int_{x_j}^{x_i}}
{\partial_x}\phi}}$, which are obtained from
those of a single chiral boson by imagining that
the $3$-lead dot/anti-dot setup is deformed as in figure 5.
As a result of the chiral boson commutations relations,
the tunneling operators commute since
they don't cross (see figure 5).
This holds whether the objects which
tunnel are fermions or anyons.

In our model, we represent ${T_{ij}}$ by
${T_{ij}} = {\eta_j}{S^\pm}{e^{\mp i{\phi_j}}}$
where $\eta_j$ and $S^\pm$ commute with each other
and with $\phi_j$, and the ${\phi_j}$'s are mutually
commuting. To ensure that the tunneling operators commute,
we must take ${\eta_i}{\eta_j}=-{\eta_j}{\eta_i}$.

As an aside, we note that if the tunneling
paths were to cross, however, the commutation relations
are modified to
${T_{ij}}{T_{kl}}+{e^{2\pi i/g}}{T_{kl}}{T_{ij}}=0$.
It is hard to imagine a setup in which this occurs, but
for such a scenario, we would need to take
the even more exotic condition
${\eta_1}{\eta_2}=-{e^{2\pi i/g}}{\eta_2}{\eta_1}$
and cyclic permutations.

\section{Instanton Gas Berry's phase Calculation}

As we briefly sketched in section III, when $t$ is large,
the interaction term dominates the action (\ref{lqbm}).
If we treat $k$ classically, it will be localized at
one of the minima of this term. To find these minima,
we need to diagonalize the $4\times 4$ matrix
${\sum_i}\left( {\eta_j}{S^+}
{e^{-i{{\bf R}^i} \cdot {\bf k} }}
\,+\, {\eta_j}{S^-} {e^{i{{\bf R}^i}
\cdot {\bf k} }}\right)$. There are $4$ solutions for
each ${\bf k}$, corresponding to the $4$ bands of a particle
on a honeycomb lattice with $\pi$ flux per plaquette.
Physically, the fourfold multiplicity is due to the
two charge states of the resonant level and the
two states of the auxiliary two-state system (i.e. $\eta$)
which keeps track of the statistics, while
${\bf k}$ represents the amount of charge which
has been transfered between the leads. Diagonalizing,
${\sum_i}\left( {\eta_j}{S^+}
{e^{-i{{\bf R}^i} \cdot {\bf k} }}
\,+\, {\eta_j}{S^-} {e^{i{{\bf R}^i}
\cdot {\bf k} }}\right)$ (we represent the $\eta_i$'s by Pauli
matrices $\tau_i$), we find the four eigenvalues:
\begin{equation}
\epsilon(k) = \pm\,{\left(3 \pm \sqrt{9 - (3 + 2\cos {k_2}/\sqrt{2} -
2\cos({k_1}\sqrt{3/2g}+{k_2}/\sqrt{2g}) -
2\cos({k_1}\sqrt{3/2g}-{k_2}/\sqrt{2g})}\right)^{1/2}}
\end{equation}
The minima of each of these bands form a honeycomb lattice
with translation vectors $2\pi{Q_i}$, where the $Q_i$
are given in (\ref{qs}).

We now consider the instanton gas expansion of the
partition function, where the instantons are solutions
of the classical equations of motion in which
${\bf k}$ tunnels between neighboring minima.
The modulus of the amplitude for these
tunneling events can be obtained in
the standard way\cite{duality}. The
phase can be obtained from the following Berry's phase argument.
The eigenvector associated with the lowest energy
band is determined by two spinors, i.e. it lies
in the direct product space of the two
two-dimensional spaces acted on by $S^{\pm}$ and
$\eta_i$. As ${\bf k}$ tunnels from minimum to
minimum, around a plaquette, these two spinors
rotate. The phase aquired in a circuit around a plaquette
is determined by the angles traced out by these
spinors, ${e^{i\phi_{\rm Berry}}} =
{e^{i({\theta_S} + {\theta_\eta})/2}}$, where the
factor of $1/2$ follows from the fact that $S$ and $\eta$
are spin-$1/2$ degrees of freedom. Since, for any
circuit, ${\theta_S}$ and
${\theta_\eta}$ must be multiples of $2\pi$,
the only possible non-trivial phase is $\pi$.

Let's consider the plaquette formed by the following six minima:
$(0,\pm \pi\sqrt{2g}/3)$, $(\pi\sqrt{2g/3},\pm 2\pi\sqrt{2g}/3)$,
$(2\pi\sqrt{2g/3},\pm \pi\sqrt{2g}/3)$. These six minima
surround a maximum of the lowest band, at $(\pi\sqrt{2g/3},0)$,
where the two lowest bands touch. The Berry phase will be
the same for any loop which encloses $(\pi\sqrt{2g/3},0)$ precisely
once since such loops can be adiabatically deformed into
each other. The Berry phase is most simply computed
for an infinitesimal loop enclosing $(\pi\sqrt{2g/3},0)$.
For such a loop, we can approximate
${\bf k} = (\pi\sqrt{2g/3},0) + {\bf p}$, and
\begin{eqnarray}
\label{diracham}
{\sum_i}\left( {\eta_j}\otimes{S^+}
{e^{-i{{\bf R}^i} \cdot {\bf k} }}
\,+\, {\eta_j}\otimes{S^-} {e^{i{{\bf R}^i}
\cdot {\bf k} }}\right)&\approx &\,\,
\left(\frac{1}{2}{S_x}-\frac{\sqrt{3}}{2}{S_y}\right)\otimes 1\,+
\,\left(\frac{\sqrt{3}}{2}{S_x}+
\frac{1}{2}{S_y}\right)\otimes\left({\eta_z} + {\eta_x}\right)\,\,-\cr
& &\,{p_x}\,\left(\frac{\sqrt{3}}{2}{S_x}+
\frac{1}{2}{S_y}\right)\otimes 1\,\,\, +\cr
& &\left(\frac{1}{2}{S_x}-\frac{\sqrt{3}}{2}{S_y}\right)
\otimes\left(\left( \frac{\sqrt{3}}{2}{p_y}-
\frac{1}{2}{p_x}\right){\eta_z} - 
\left( \frac{\sqrt{3}}{2}{p_y}+
\frac{1}{2}{p_x}\right){\eta_z}
\right)
\end{eqnarray}
As ${\bf p}$ adiabatically traces out a loop enclosing
${\bf p}=0$, the effective Zeeman field `seen' by
${\bf \eta}$ traces out a circle but the effective
Zeeman field `seen' by ${\bf S}$ does not. In other
words ${\theta_\eta}=2\pi$ while ${\theta_S} = 0$.
This can be made more transparent by projecting (\ref{diracham})
onto the two-dimensional subspace which is degenerate
at ${\bf p}=0$. In an orthonormal basis of the
two eigenvectors with degenerate eigenvalues at
${\bf p}=0$, (\ref{diracham}) can be rewritten as:
\begin{equation}
h({p_x},{p_y})\cdot{\bf \tau}\,
+ \,{\rm const.}
\end{equation}
where $\bf \tau$ are a set of Pauli matrices
and $h({p_x},{p_y})$ rotates by $2\pi$ as ${\bf p}$ adiabatically
traces out a loop enclosing ${\bf p}=0$.

Hence, a $\pi$ phase is aquired in a circuit about a plaquette.
Combining this with the magnitudes of the terms
in the standard Coulomb gas expansion
for the instanton gas \cite{duality},
we see that the dual theory to (\ref{lqbm}) is also
a theory defined on a honeycomb lattice with $\pi$-flux,
namely (\ref{ldual}).

\section{ Boundary conditions on bosons}

In this appendix we summarize the conventions which we use
for compactified bosons.
Consider a single boson $\varphi(x,\tau)$ compactified
on a circle of radius $r$.  On a space of size $l$ with
periodic boundary conditions the action is
$$ S_0
=  
{1\over 8 \pi}
\int_{0}^{l} dx
\int_{-\infty}^{\infty} d \tau
(\partial_{\mu}\varphi) (\partial^{\mu} \varphi)
$$
where the functional integral is to be performed under the
identification
$$
 \varphi(x,\tau) =
 \varphi(x,\tau)  + 2 \pi r
 =\varphi(x+l,\tau) 
$$
In the Hamiltonian formalism, the field operator is
$$
\varphi(x)
=
\varphi_L(x)
+
\varphi_R(x)
$$
where
$$
\varphi_L(x)
= X_L + x \ P_L {2\pi \over l} +
 i \sum_{n=1}^{\infty} {1\over \sqrt{n} }
[ b_{L;n} e^{i x n 2 \pi/l} + b^{\dagger}_{L;n} e^{-i x n  2 \pi/l} ]
$$
\begin{equation}
\varphi_R(x)
= X_R - x \ P_R {2\pi \over l} +
i \sum_{n=1}^{\infty} {1\over \sqrt{n} }
[ b_{R;n} e^{-i x n  2 \pi/l} + b^{\dagger}_{R;n} e^{i x n  2 \pi/l} ]
\label{bosonfieldoperator}
\end{equation}

$b_{L;n}$ and $b_{R;n}$ are independent boson creation and annihilation
operators, and
$$
[X_L,P_L]=
[X_R,P_R]=i ({l \over 2 \pi })
$$
are two independent zero mode coordinate and momentum operators.
In these conventions, the two point funtion of the
boson is
\begin{equation}
<\varphi(x,\tau) \varphi(0,0)>
=
- \ln|z|^2,
\qquad (z=\tau+i x )
\label{twopointboson}
\end{equation}

The momentum operator, conjugate to the field $\varphi(x)$
is 
\begin{equation}
\Pi(x)
=
(1/2) 
{2 \pi   \over l}
\bigl \{ 
(P_L + P_R)
+
\sum_{n=1}^{\infty}
 \sqrt{n} 
[
b_{L;n}
e^{i x n  2 \pi/l}
 +
b^{\dagger}_{L;n}
e^{-i x n 2 \pi/l}
]
+
\sum_{n=1}^{\infty}
 \sqrt{n} 
[
b_{R;n}
e^{-i x n  2 \pi/l}
 +
b^{\dagger}_{R;n}
e^{i x  n 2 \pi/l}
]
\bigr \}
\label{bosonmomentum}
\end{equation}
The total momentum
$$
\Pi = \int_{0}^l d x \Pi (x) = (1/2)[ P_L + P_R] {2 \pi  \over l}
$$
is canonical conjugate to the total zero mode coordinate $X \equiv X_L + X_R$.
Since the latter is  periodic with period $2 \pi r$,
the eigenvalues of  `dimensionless momentum' $P$ must be of the form
\begin{equation}
 P = [ P_L + P_R] = { 2n \over r}, \qquad n\in Z
\label{Pquantization}
\end{equation}

On the other hand, periodicity under $x \to x + l$ gives, using
(\ref{bosonfieldoperator})
\begin{equation}
W\equiv P_L- P_R =  r m, \qquad m \in Z
\label{windingquantization}
\end{equation}
where we denote this quantity by the {\it winding number}
$W$.
The two conditions (\ref{Pquantization}) and
(\ref{windingquantization}) imply together that

\begin{equation}
P_L = {n\over r} + {1\over 2} r m;
\qquad
P_R = {n\over r} - {1\over 2} r m;
\label{windingmomentum}
\end{equation}

The (normal ordered)  hamiltonian is
\begin{equation}
H ={2 \pi \over l}[ L_{0;L} +L_{0;R}  - {1\over 12}]
\label{hamiltoniancompact}
\end{equation}
where
\begin{equation}
 L_{0;L}=
{1\over 2} P_L^2 +
\sum_{n=1}^{\infty}
  \ n \ b^{\dagger}_{L;n}
b_{L;n},
\qquad
 L_{0;R}=
{1\over 2} P_R^2 +
\sum_{n=1}^{\infty}
 \ n \ b^{\dagger}_{R;n}
b_{R;n}
\label{Lnot}
\end{equation}
Note that 
$$
{1\over 2}[ P_L^2 +P_R^2 ]=[(n/r)^2 + (rm/2)^2]
\label{scalingdims}
$$ 
gives
the total scaling dimension.

The (imaginary time evolved) Heisenberg operators are obtained
from the expressions in 
( \ref{bosonfieldoperator},\ref{bosonmomentum})
by $ix \to  z=\tau+ix $ for left movers, and
by  $ix \to  z^*=\tau-ix $ for right movers.

Of interest is also the {\it dual} boson field,
\begin{equation}
{\tilde \varphi}(x)
=
\varphi_L(x) - \varphi_R(x)
\label{dualbosonfieldoperator}
\end{equation}
We see that the  automorphism of the canonical commutation relations
\begin{equation}
X_R \to  -X_R, 
\quad
P_R \to - P_R,
\quad
b_{R,n} \to
-
b_{R,n}
\quad
{b^\dagger_{R,n}} \to
-
{b^\dagger_{R,n}},
\qquad
r \to {\tilde r} \equiv 2/r
\label{duality}
\end{equation}
(all left-movers unchanged)
maps the boson field $\varphi(x)$ into its dual ${\tilde \varphi}(x)$.
The dual field then satisfies the periodicity condition
$
{\tilde \varphi}(x)
=
{\tilde \varphi}(x)
+
 2\pi {\tilde r}
$ where
 $\tilde r$
is the 
{\it dual} compactification radius.
Note that, most importantly, the duality automorphism
exchanges the lattice of momenta $P$ (\ref{Pquantization})
with the lattice of winding numbers  $W$  (\ref{windingquantization}).

\vskip .5cm
\noindent{\it Boundary Conditions:}
Next, consider the compactified boson in {\it semi-infinite}
space, $0<x<\infty$. First, we impose a Neumann  boundary condition (b.c.)
on the  boson field at $x=0$,
\begin{equation}
\partial_x \varphi(x,\tau) \to 0 \qquad (x\to 0)
\label{Neumannbc}
\end{equation}

For the purpose of analyzing this b.c. it is convenient to
view the imaginary time coordinate as a spatial coordinate
${\tilde x}\equiv \tau$, and the original spatial coordinate $x$ as (imaginary)
time, ${\tilde \tau} \equiv -x$.
The new complex coordinates, are just rotated by $90$ degrees
with respect to the original ones,
${\tilde z} \equiv {\tilde \tau} + i {\tilde x}= i z $,
${\tilde z}^* \equiv {\tilde \tau} - i {\tilde x}= (-i )z^* $
 (this may be viewed as a trivial conformal transformation).
We may quantize the system as before, but now on equal $\tilde \tau$
slizes. The 'spatial` coordinate ${\tilde x}=\tau$ is now compact,
corresponding to the original system being at finite inverse
temperature $\beta  < \infty$. 
The Heisenberg operators in this quantization become

$$
\varphi_L({\tilde x},{\tilde \tau})
= X_L  -i({\tilde \tau} + i {\tilde x}) \ P_L  
{2 \pi \over \beta}
+
 i \sum_{n=1}^{\infty} {1\over \sqrt{n} } 
[ b_{L;n} e^{ ({\tilde \tau} + i{\tilde  x})  n 2 \pi/\beta} +
b^{\dagger}_{L;n} e^{-({\tilde \tau} + i{\tilde  x}) n  2 \pi/\beta} ]
$$
\begin{equation}
\varphi_R({\tilde x},{\tilde \tau})
= X_R -i ({\tilde \tau} - i{\tilde  x}) \ P_R 
{2 \pi \over \beta}
+
i \sum_{n=1}^{\infty} {1\over \sqrt{n} }
[ b_{R;n} e^{ ({\tilde \tau} - i{\tilde  x}) n  2 \pi/\beta}
 + b^{\dagger}_{R;n} e^{ -({\tilde \tau} - i{\tilde  x}) n  2 \pi/\beta} ]
\label{modularbosonfieldoperator}
\end{equation}

Note that the currents associated with translations of the two
chiral bosons are the left/right momenta:
$$
J_L = i \partial_{\tilde z} \varphi_L = P_L {2 \pi \over \beta} + oscillators,
\qquad
J_R = i \partial_{\tilde  {\bar z}} \varphi_R = P_R {2 \pi \over \beta}
+ oscillators
$$

The Neumann b.c. now becomes an identity for Heisenberg operators,
acting on a boundary state $|N>$:
\begin{equation}
\partial_{\tilde \tau} {\hat \varphi}({\tilde x}, {\tilde \tau})|N> \to 0 
\qquad ({\tilde \tau}\to 0)
\label{neumannstate}
\end{equation}
Integrating this equation over $\tilde x$
implies in particular that  the total momentum operator
${\hat P}= {\hat P}_L+{\hat P}_R$ annihilates
the Neumann boundary state 
 but 
the winding number $W= P_L - P_R = 2 P_L $ may take on any value
on the lattice
specified in  (\ref{windingquantization}).
(This is clear since
the Neuman b.c. only constrains the derivative of the
field.) 
Specifically, the only momentum states which
the Neumann boundary condition supports are those
with  $n=0$ in (\ref{windingmomentum}).
Unfolding the semi-infinite system with a boundary
thus gives an infinite chiral system of bosons
with only those momenta allowed.

Next consider the Dirichlet boundary condition
$$
\varphi(x,\tau) \to 0 \qquad (x\to 0)
$$
After $90^o$ degree rotation, this becomes an identity
for Heisenberg operators, acting on a `Dirichlet' boundary
state $|D>$:
$$
{\hat \varphi}({\tilde x}, {\tilde \tau}) |D> \to 0
\qquad ({\tilde \tau}\to 0)
$$
This implies
that the total winding number operator, ${\hat W} = {\hat P}_L - {\hat P}_R $
 annihilates the Dirichlet boundary state,
but the total momentum $P=2P_L$ lies on the lattice
specified in (\ref{Pquantization}).
(The winding number is seen to be  zero also because  of the
operator identity $X_L + X_R =0$  which follows from the Dirichlet
b.c..)
Specifically, the only momentum states which
the Dirichlet boundary condition supports are those
with  $m=0$ in (\ref{windingmomentum}).
Unfolding the semi-infinite systems gives thus
an infinite  chiral system with only those momenta allowed.

\noindent {\it Duality and boundary conditions.}
It immediately follows from the discussion in the paragraph
above that the duality operation exchanges Dirichlet
and Neuman boundary conditions. In other words,
a Dirichlet b.c.  on the field $\varphi$ 
is a Neuman b.c. on the dual field ${\tilde \varphi }$
and vice versa.

\section{Transformation of the Conductance under Duality}

The conductance can be  obtained from the
two point function

$$
<J^3(\tau_1,x_1)J^3(\tau_2,x_2)>
$$
where
$$
J^3(\tau, x) = J^3_L + J^3_R 
$$
and $J^3_{R,L}$ is proportional to $\partial \varphi^3_{R,L}$.
At the location of the point contact,
we obtain four terms, which are pairwise equal
$$
<J^3(\tau_1,0)J^3(\tau_2,0)> =
2(1+A)/{({\tau_1}-{\tau_2})^2}
$$
where
$A$ is the amplitude of the
right-left current-current correlator,
$$
<J^3_L(\tau_1,x_1)J^3_R(\tau_2,x_2)>
=
A/(z_1-z^*_2)^2
$$
($z_i = \tau_i + i x_i$),
and 
$$
<J^3_L(\tau_1,x_1)J^3_L(\tau_2,x_2)>
=
1/z_{12}^2
$$
(similar for right movers).

The conductance can be written as
$$
G
=
g( 1 - A)
$$ 

This we may rewrite as
$$
G= 2g[ 1/2 - A/2]=
2g[ 1 - (1+A)/2]
$$
On the other hand we have
$$
(1+A)/2
=
{|\omega| \over 2 \pi} <\varphi^3(x=0,\omega) \varphi^3(x=0,\omega) >
$$
where $\omega$ is a real frequency,
and $\varphi^3 = \varphi^3_L + \varphi^3_R$.
Under duality, $A \to -A$, which may also be written as
$$
{|\omega| \over 2 \pi} <\varphi^3(x=0,\omega) \varphi^3(x=0,\omega) >
\to (1 
-
{|\omega| \over 2 \pi}
 <{\tilde \varphi}^3(x=0,\omega)  {\tilde \varphi}^3(x=0, \omega) >)
 $$
Clearly, for Neumann boundary conditions, we have have $A=1$,
yielding $G=0$. For Dirichlet boundary condition on
$\varphi^3$ (as opposed to Dirichlet boundary conditions
on $k_x$, $k_y$), $A=-1$,
yielding, $G=2g$.

Since we will not impose boundary conditions on the fields $\varphi^j$
directly, we rewrite the conductance in terms of the fields
defined in Eq(2.7), giving
$$
G=
{ 4g \over 3}
[1 -  {|\omega| \over 2 \pi}<k_x k_x>_{\omega}]
$$

Under the duality transformation, $k_x \to {\tilde k_x} = r_x$,
we have:
$$
G \to {\tilde G}
={4 {\tilde g}\over 3}
{|\omega| \over 2 \pi}<r_x r_x>_{\omega}
$$
At the selfdual point, $g = {\tilde g} =\sqrt{3}$,
and 
$[1 -  {|\omega| \over 2 \pi}<k_x k_x>_{\omega}]=
{|\omega| \over 2 \pi}<r_x r_x>_{\omega}
=1/2$.

\newpage

\begin{figure}
\centerline{\psfig{figure=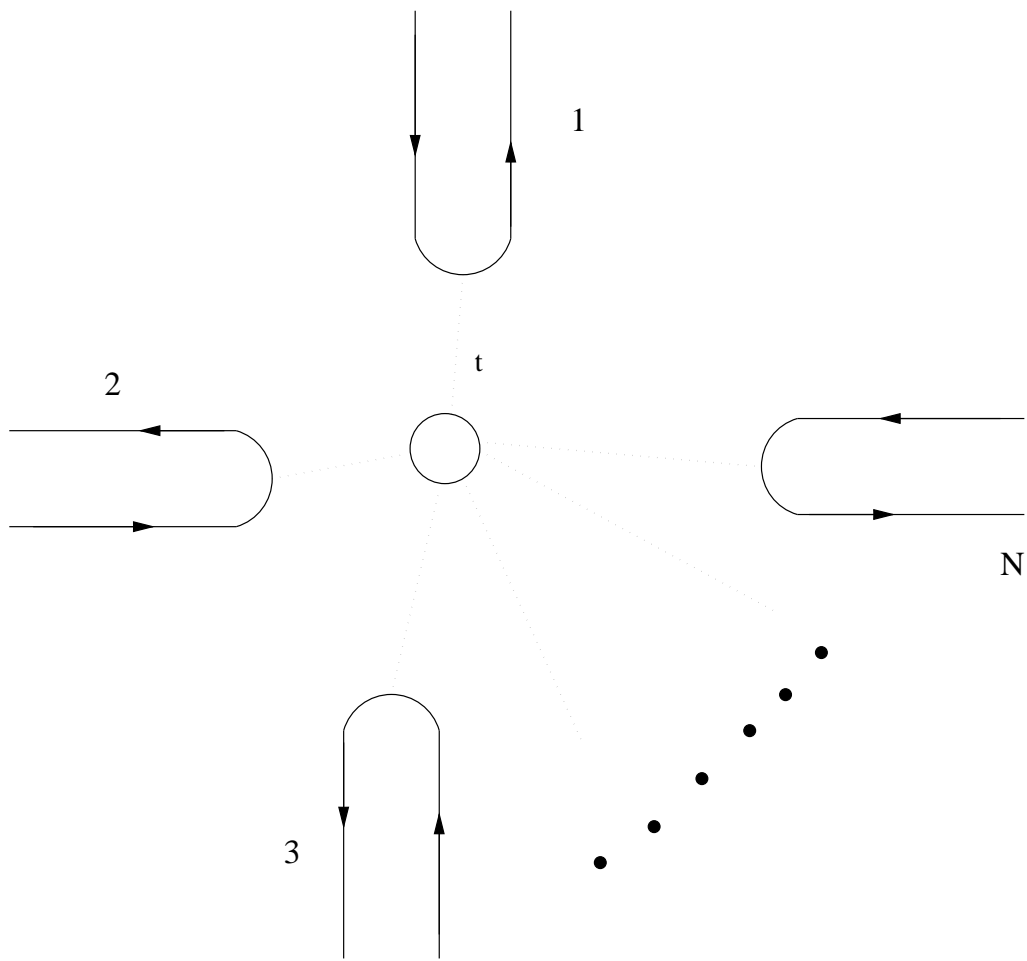,height=5in,angle=-90}}
\vskip 0.5cm
\caption{A multi-lead resonant tunneling setup.}
\end{figure}

\newpage

\begin{figure}
\centerline{\psfig{figure=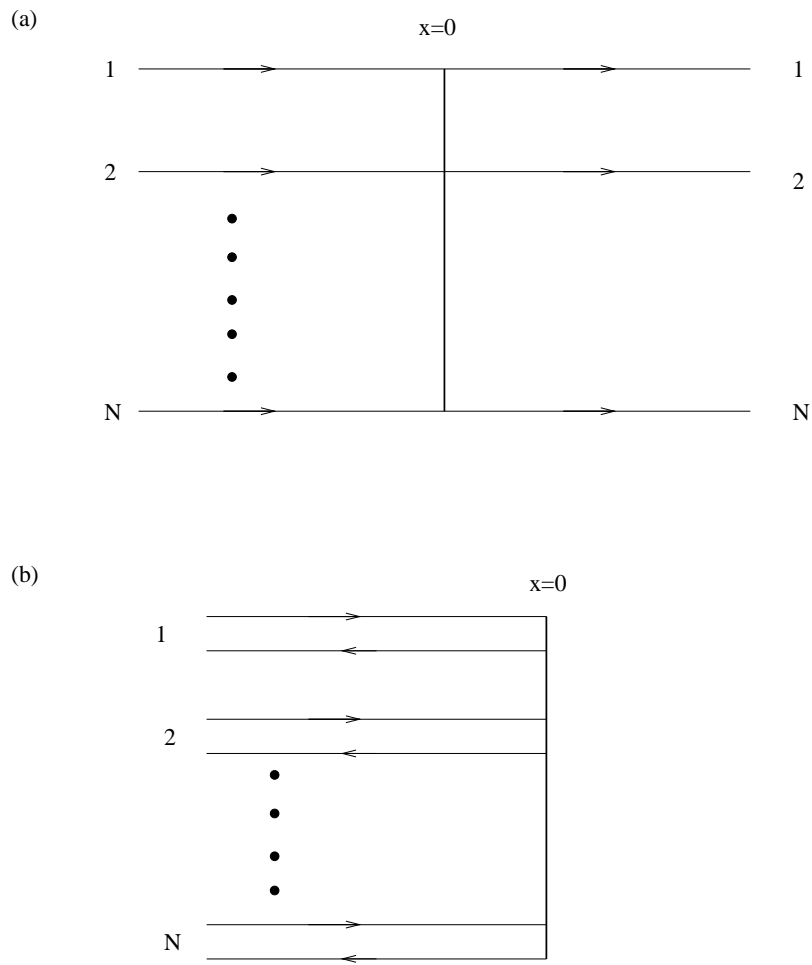,height=5in,angle=-90}}
\vskip 0.5cm
\caption{The physical picture in the (a) `unfolded' formalism
and (b) `folded' formalism.}
\end{figure}

\newpage

\begin{figure}
\centerline{\psfig{figure=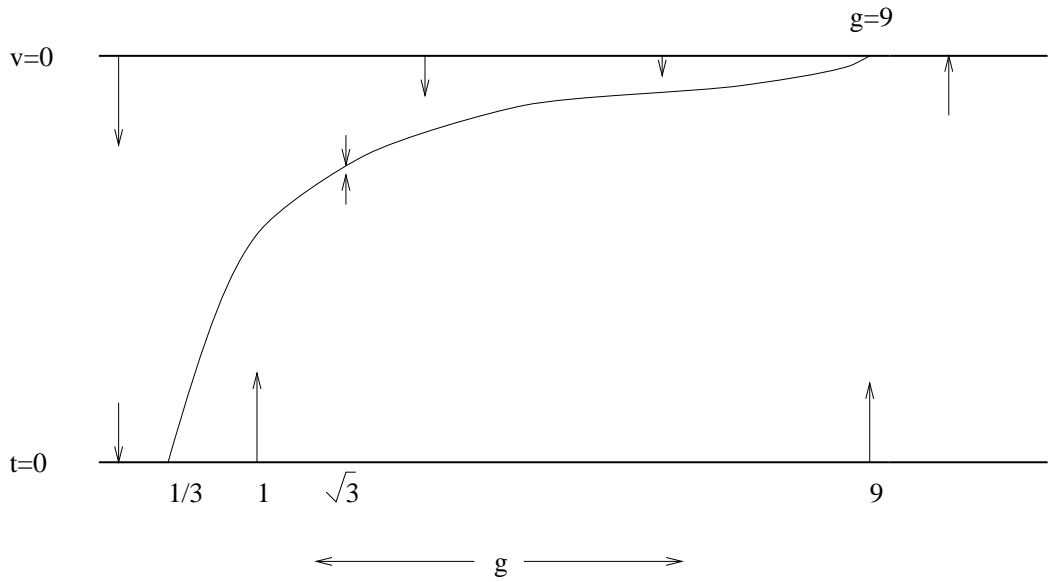,height=3in,angle=-90}}
\vskip 0.5cm
\caption{The phase diagram. The horizontal axis measures $g$,
the Luttinger liquid parameter, and the vertical axis measures
$t$, the hopping strength, and its dual variable, $v$. The
RG flows are as indicated. The intermediate coupling fixed
points which are stable for $1/3<g<9$ are represented
by the curve connecting the weak and strong-coupling
fixed points at $g=1/3$ and $g=9$, respectively.}
\end{figure}

\newpage

\begin{figure}
\centerline{\psfig{figure=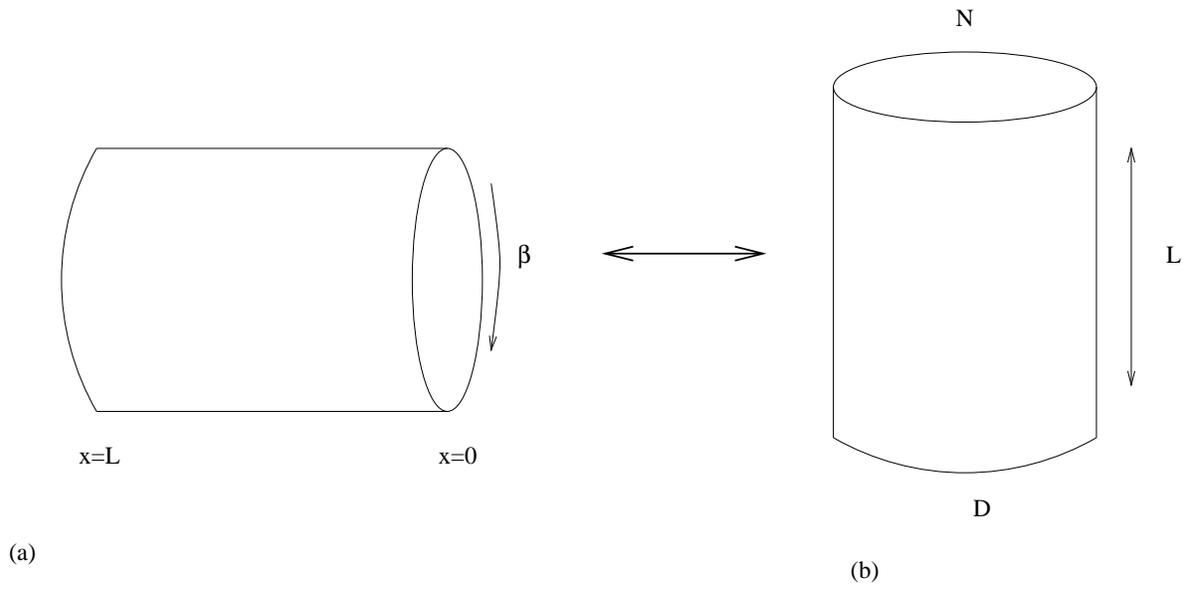,height=3in,angle=-90}}
\vskip 0.5cm
\caption{The (a) finite-temperature partition function
of this model can be represented as (b) the closed
string amplitude for propagation between boundary states
$D$ and $N$.}
\end{figure}

\newpage

\begin{figure}
\centerline{\psfig{figure=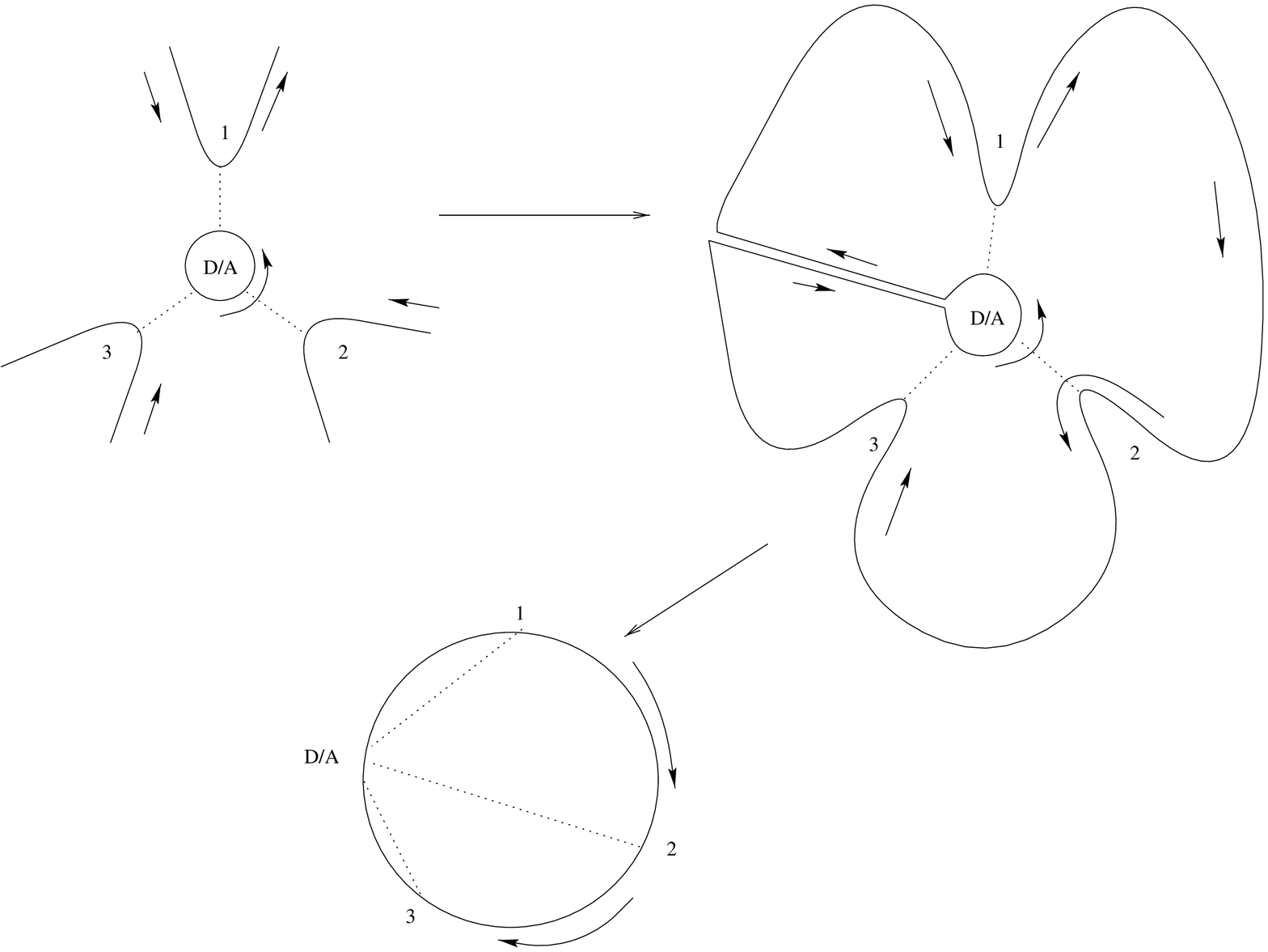,height=5in,angle=-90}}
\vskip 0.5cm
\caption{Deformation of the resonant tunneling arrangement
into a chiral boson on a circle.}
\end{figure}


\begin{thebibliography}{99}


\bibitem{affleckludwig} see e.g.:
 I Affleck, A.W.W.Ludwig,  Nucl. Phys. B360
(1991) 641; Nucl. Phys. B428 (1994) 545;
  A.W.W.Ludwig, Int. J. Mod. Phys. B8 (1994) 347; 
I.Affleck, Acta Phys. Polon. B26 (1995) 1869;
S.Eggert, I .Affleck, Phys. Rev. B46 (1992) 10866.

\bibitem{bethe}  P.B.Wiegman , A.M.Tsvelik, JETP Lett. 38 (1983), 591;
Adv. Phys. 32 (1983) 543; N.Andrei, K.Furuya, J.Lowenstein,
Rev. Mod. Phys. 55 (1983) 331.

\bibitem{FLS}  P.Fendley, A.W.W.Ludwig, H.Saleur
Phys. Rev. Let. 74 (1995) 3005; Phys Rev. B 52 (1995) 8934


\bibitem{yikane} H. Yi and C.L. Kane, cond-mat/9602099.

\bibitem{kanefisher} C.L. Kane and M.P.A. Fisher,
Phys. Rev. Lett. {\bf 68}, 1221 (1992); Phys. Rev.
{\bf B 46}, 15233 (1992).

\bibitem{emerykiv} V.J. Emery and S. Kivelson,
Phys. Rev. {\bf B 46}, 10812 (1992).

\bibitem{duality} A. Schmid, Phys. Rev. Lett. {\bf 51}, 1506 (1983);
M.P.A. Fisher and W. Zwerger, Phys. Rev. {\bf B 32}, 6190 (1985).

\bibitem{sengupta} A.M. Sengupta, Y.B.Kim, cond-mat/9602100.

\bibitem{chamonfradkin}C. de C. Chamon and E. Fradkin,
Phys. Rev. {\bf B 56}, 2012 (1997); N.P. Sandler,
C. de C. Chamon, and E. Fradkin, cond-mat/9704189.
 

\bibitem{CRE} For a discussion using a language related to this
paper see: J Polchinski, Nucl. Phys. B242 (1984) 345; I.Affleck
and J Sagi, Nucl. Phys. B417 (1994) 374.


\end{thebibliography}
\end{document}